\documentclass[twocolumn]{revtex4-1}
\usepackage{amsmath}
\usepackage{mathtools}

\begin{document}
\title{A light weight regularization for wave function parameter gradients
\\ in quantum Monte Carlo}

\author{Shivesh Pathak}
\email[]{sapatha2@illinois.edu}

\author{Lucas K. Wagner}
\email[]{lkwagner@illinois.edu}
\affiliation{Department of Physics; University of Illinois at Urbana-Champaign}

\date{\today}
\begin{abstract}
The parameter derivative of the expectation value of the energy, $\partial E/\partial p$, is a key ingredient in variational quantum Monte Carlo (VMC) wave function optimization methods.
In some cases, a na\"ive Monte Carlo estimate of this derivative suffers from an infinite variance which inhibits the efficiency of optimization methods that rely on a stable estimate of the derivative.
In this work, we derive a simple regularization of the na\"ive estimator which is trivial to implement in existing VMC codes, has finite variance, and a negligible bias which can be extrapolated to zero bias with no extra cost.
We use this estimator to construct an unbiased, finite variance estimation of $\partial E/\partial p$ for a multi-Slater-Jastrow trial wave function on the LiH molecule.
This regularized estimator is a simple and efficient estimator of $\partial E/\partial p$ for VMC optimization techniques.
\end{abstract}
\maketitle 

\section{Introduction}

Variational Monte Carlo (VMC) is a powerful technique to study correlated many-electron wave functions in materials.
Accurate and stable stochastic estimation of the parameter gradient of the energy expectation value is an integral component of VMC wave function energy optimization techniques \cite{PhysRevB.64.024512, doi:10.1063/1.1604379, Toulouse2007, Umrigar2005, Umrigar2007, Toulouse2008}.
Within these techniques, the parameters $\vec{p}$ in a wavefunction $\Psi(\vec{R},\vec{p})$ are optimized to minimize the energy expectation value $E(\vec{p}) \equiv \langle \Psi(\vec{p})|\hat{H} |\Psi(\vec{p})\rangle$.
This is accomplished through iterative approaches where a stochastic estimate of the gradient $\partial E(\vec{p})/\partial p$ is used in updating the parameter $p$ at each iteration. 
Low bias and low variance evaluations of the derivative directly improve the efficiency of stochastic wave function optimization techniques. 

The simplest estimator for $\partial E/\partial p$ is the derivative of the energy expectation value (Eqn~\ref{eq:naive_estimator} in the methods).
This estimator is unbiased and has a zero-variance principle for any wave function parameter $p$ if $\Psi(\vec{R}, \vec{p})$ is an eigenstate of $\hat{H}$.
However, the variance of this estimator diverges when $\Psi$ is not an exact eigenstate and $p$ is a parameter which affects the nodes of the wave function, as occurs in the case of orbital or determinant coefficients \cite{Avella, doi:10.1063/1.4933112}.

Previous to this work, the infinite variance has been removed in several ways. 
One method is to sample from distributions other than the trial wave function squared.\cite{Avella, Attaccalite2008, Zen2013} 
This method has a small bias\cite{doi:10.1063/1.4933112} due to nonlinearity in the estimator, which is removed as the sample size increases.
Alternatively, one can modify the estimator using an auxillary wave function.\cite{Assaraf1999, doi:10.1063/1.1286598, Assaraf2003}
However, both of these approaches require a choice of guiding or auxillary functions to implement the algorithm.

In this paper, we derive and test a simple regularized estimator for $\partial E/\partial p$ which has finite variance and can be efficiently extrapolated to zero bias.
Instead of guiding wave functions, the divergent variance of the na\"ive Monte Carlo estimator is removed via multiplication by a polynomial function within a distance $\epsilon$ of the nodes of $\Psi(\vec{p}, \vec{R})$. 
We show that the bias scales as $\epsilon^3$ and the variance scales as $1/\epsilon$. 
This estimator requires essentially no modification to simple VMC codes. 
For a realistic system, we test that the predicted scaling is seen in a real calculation, and show that extrapolation to zero bias is straightforward and requires no extra computation.

\section{Regularized estimator}
The na\"ive Monte Carlo estimate for $\partial E/\partial p$ evaluated on a wave function $\Psi(\vec{R}, \vec{p})$ is 
\begin{equation}
\hat{\theta} \equiv \left\langle \Big(E_L(\vec{R})  - \left\langle E_L(\vec{R}) \right \rangle\Big)\frac{\partial_p \Psi(\vec{R}, \vec{p})}{\Psi(\vec{R}, \vec{p})} \right\rangle
\label{eq:naive_estimator}
\end{equation} 
where the brackets $\langle \ \rangle$ are Monte Carlo expectation values over $M$ configurations drawn from the distribution $|\Psi(\vec{p}, \vec{R})|^2$, and $E_L = (\hat{H}\Psi)/\Psi$ is the local energy with Hamiltonian operator $\hat{H}$.
This estimator has zero bias for any wave function $\Psi$ and parameter $p$, but can acquire a divergent variance.
The divergent contribution to the variance arises from the evaluation of the integral
\begin{equation}
\int \Big(E_L\frac{\partial_p\Psi}{\Psi}\Big)^2 |\Psi|^2 dR.
\label{eq:divergent_integral}
\end{equation}
when $\Psi$ is not an eigenstate of $\hat{H}$ and the parameter $p$ affects the nodes of $\Psi$, such as orbital or determinantal coefficients.
In this case, to lowest order in distance $|\vec{R}-\vec{N}|$ from a nodal point $\vec{N}$ of $\Psi$, $\hat{H}\Psi$ and $\partial_p \Psi \simeq const$ and $\Psi \simeq \nabla \Psi(\vec{N}) \cdot (\vec{R} - \vec{N})$, leading to the integrand in Eqn~\ref{eq:divergent_integral} behaving as $1/|\vec{R}-\vec{N}|^2$ near the node.
Since the integration domain includes the nodal surface of $\Psi$, the $1/|\vec{R}-\vec{N}|^2$ behavior of the integrand as $|\vec{R}-\vec{N}|\rightarrow 0$ leads to a divergent integral.

We obtain a finite variance estimator for $\partial E/\partial p$ by regularizing the na\"ive estimator of Eqn~\ref{eq:naive_estimator} by a function $f_\epsilon$:
\begin{equation}
\hat{\theta}_\epsilon \equiv \left\langle \Big(E_L(\vec{R})  - \left\langle E_L(\vec{R}) \right \rangle\Big)\frac{\partial_p \Psi(\vec{R}, \vec{p})}{\Psi(\vec{R}, \vec{p})} f_\epsilon(\vec{R}) \right\rangle
\label{eq:regularized_estimator}
\end{equation}
where 
\begin{equation}
\begin{split}
&f_\epsilon(\vec{R}) = \begin{cases} 
     7(\frac{\vec{x}}{\epsilon})^6 - 15(\frac{\vec{x}}{\epsilon})^4 + 9(\frac{\vec{x}}{\epsilon})^2 & |\frac{\vec{x}}{\epsilon}| < 1 \\
      1 & |\frac{\vec{x}}{\epsilon}| \ge 1 \\
   \end{cases}\\ 
 &\vec{x} \equiv \frac{\Psi(\vec{R, \vec{p}}) \nabla \Psi(\vec{R}, \vec{p})}{|\nabla \Psi(\vec{R}, \vec{p})|^2}.
\end{split}
\label{eq:regularizing_function}
\end{equation} 
The quantity $|\vec{x}|$ is the normal distance between $\vec{R}$ and nearest nodal point of $\Psi$, $\epsilon$ is a parameter which defines how far away from the node the regularization should be carried out, and $\nabla$ is the many-body gradient.
Details on the construction of $f_\epsilon$ can be found in the Appendix.

In evaluating the variance of Eqn~\ref{eq:regularized_estimator}, the divergent integral Eqn~\ref{eq:divergent_integral} is replaced by a regularized integral
\begin{equation}
\begin{split}
\int \Big(E_L\frac{\partial_p\Psi}{\Psi}\Big)^2 f_\epsilon^2 |\Psi|^2 dR.
\end{split}
\label{eq:convergent_integral}
\end{equation}
Unlike the integrand of Eqn~\ref{eq:divergent_integral}, the regularized integrand does not diverge near the nodes of $\Psi$. 
This is seen by Taylor expanding the integrand to lowest non-vanishing order in $|\vec{R} - \vec{N}|$. 
Substituting the limiting relationships for $\hat{H}\Psi, \ \partial_p \Psi$ and $\Psi$ from above and noting 
\begin{equation}
   \vec{x} \simeq \Big[\nabla\Psi(\vec{N}) \cdot (\vec{R} - \vec{N}) \Big] \frac{\nabla \Psi(\vec{N})}{|\nabla \Psi(\vec{N})|^2}
\end{equation} 
 as $|\vec{R} - \vec{N}| \rightarrow 0$, we find the limiting behavior of the regularized integrand near the nodes of $\Psi$: 
\begin{equation}
\lim_{|\vec{R} - \vec{N}| \rightarrow 0 } \Big(E_L\frac{\partial_p\Psi}{\Psi}\Big)^2 f_\epsilon^2 |\Psi|^2 \propto \frac{|\nabla\Psi(\vec{N}) \cdot (\vec{R} - \vec{N})|^2}{|\nabla\Psi(\vec{N)}|^4 \epsilon^4}\rightarrow 0.
\label{eq:convergent_integrand}
\end{equation}
The regularized estimator thus removes the divergent contribution to the variance present in Eqn~\ref{eq:divergent_integral} and results in a finite variance estimation of $\partial E/\partial p$ for any $\epsilon > 0$.

To lowest order in $\epsilon$, the variance of Eqn~\ref{eq:regularized_estimator} decreases as $1/\epsilon$.
This can be seen by carrying out the integral Eqn~\ref{eq:convergent_integral} as $\epsilon\rightarrow 0 $. 
Since the regularization is only present for $|\vec{x}/\epsilon| < 1$, the lowest order scaling with $\epsilon$ appears in the integration domain $|\vec{x}/\epsilon| < 1$. 
Within this domain, the limit $\epsilon \rightarrow 0$ is equivalent to $|\vec{R} - \vec{N}| \rightarrow 0$, and the limiting integrand of Eqn~\ref{eq:convergent_integrand} can be used. 
Making the change of variables $\vec{R} \rightarrow (l, \vec{N})$ where $l$ is the normal coordinate to the nodal surface and $\vec{N}$ is the coordinate of the nodal surface, Eqn~\ref{eq:convergent_integral} to lowest order in $\epsilon$ is 
\begin{equation}
\propto \int dN \frac{1}{|\nabla\Psi(\vec{N})|^2} \int_{-\epsilon}^{\epsilon} dl \frac{l^2}{\epsilon^4} \propto \frac{1}{\epsilon}.
\end{equation}

The suppressed variance of the estimate comes at a cost, in the form of cubic bias $O(\epsilon^3)$ to lowest order in $\epsilon$.
The bias of the regularized estimator is
\begin{equation}
\int_{|\vec{x}/\epsilon|< 1} (\hat{H}\Psi) (\partial_p \Psi) (f_\epsilon - 1) dR.
\label{eq:estimator_bias}
\end{equation}
Following the analysis of the previous paragraph, we evaluate the bias to lowest non-vanishing order in $\epsilon$ by Taylor expanding the integrand in powers of $|\vec{R} - \vec{N}|$ and calculating the bias at each order in the expansion.
The constant term in the Taylor expansion does not contribute to the bias by parity of the integrand.
Further, the choice of coefficients for $f_\epsilon$ in Eq~\ref{eq:regularizing_function} ensures the first order term vanishes, as proven in the appendix.
The lowest non-vanishing contribution to the bias is due to term in the Taylor expansion $\propto l^2$, yielding an integral which scales as $\epsilon^3$.

This cubic bias can be removed by a zero-bias extrapolation carried out in four steps.
\begin{enumerate}
\item Conduct a standard VMC calculation to collect $M$ configurations $\{\vec{R}_i\}_{i=1}^M$ from $|\Psi|^2$.
\item Evaluate $|\vec{x}_i|$ from Eqn~\ref{eq:regularizing_function} for each configuration.
\item Calculate the regularized estimate Eqn~\ref{eq:regularized_estimator} for a sequence of $\epsilon \rightarrow 0$ on the collected configurations.
\item Fit a function $a\epsilon^3 + b$ to the estimates; the intercept $b$ is the zero bias, finite variance estimate for $\partial E/\partial p$.
\end{enumerate}
This procedure can be carried out very efficiently as only a single set of VMC configurations must be drawn for the entire extrapolation, resulting in an inexpensive, finite-variance, zero bias estimate of $\partial E/\partial p$.

\section{Application to LiH molecule}
We verify the predicted mathematical behavior of $\hat{\theta}_\epsilon$ by computing $\partial E/\partial p$ for a determinantal coefficient of an unoptimized multi-Slater Jastrow (MSJ) wave function for the LiH molecule:
\begin{equation}
\Psi(\vec{c}, \vec{\alpha}) = e^{J(\vec{\alpha})} \sum_{i} c_i  |D_i \rangle.
\end{equation}
The determinants $|D_i \rangle$ and coefficients $c_i$ were taken from a full configuration interaction (CI) expansion over the Li 1s, 2s, 2p and H 1s orbitals.
The orbitals were constructed from a restricted open-shell Hartree Fock (ROHF) calculation using a correlation consistent quadruple-zeta valence basis set \cite{doi:10.1063/1.456153}.
A 2-body Jastrow factor $J(\vec{\alpha})$ of the form in \cite{Wagner2009} was used with $\vec{\alpha} = 0$ except for electron-electron cusp conditions.
The ROHF and CI calculations were done using the PySCF package \cite{PYSCF} and all QMC calculations were carried out using \texttt{pyqmc}\cite{pyqmc}.

\begin{figure}
\includegraphics[width=1.0\columnwidth]{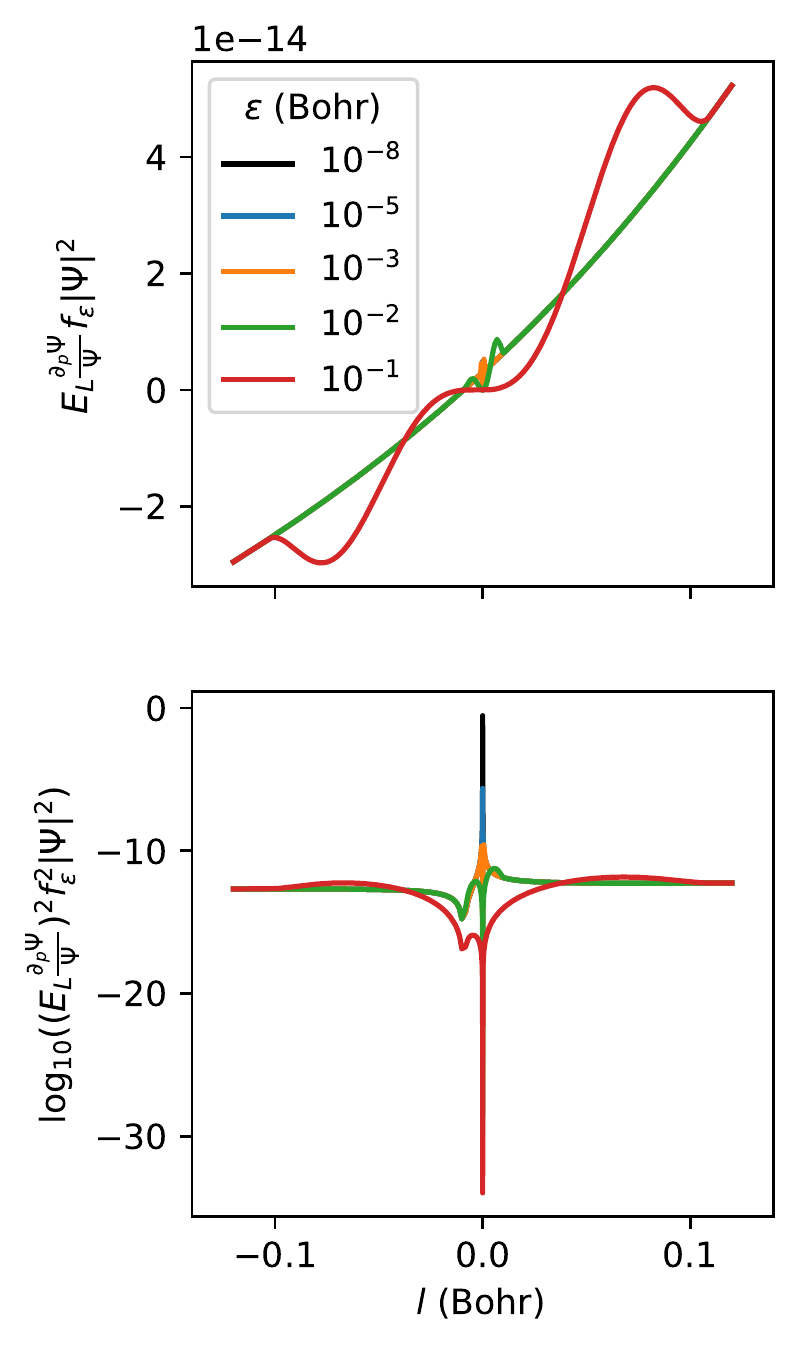}
\caption{$E_L\frac{\partial_p \Psi}{\Psi} f_\epsilon$ and logarithm $(E_L\frac{\partial_p \Psi}{\Psi} f_\epsilon)^2$, plotted against the normal coordinate $l$ from a node of $\Psi$. Curve colors correspond to different values of $\epsilon$ ranging from $10^{-1}$ to $10^{-8}$ Bohr.}
\end{figure}

\begin{figure}
\includegraphics{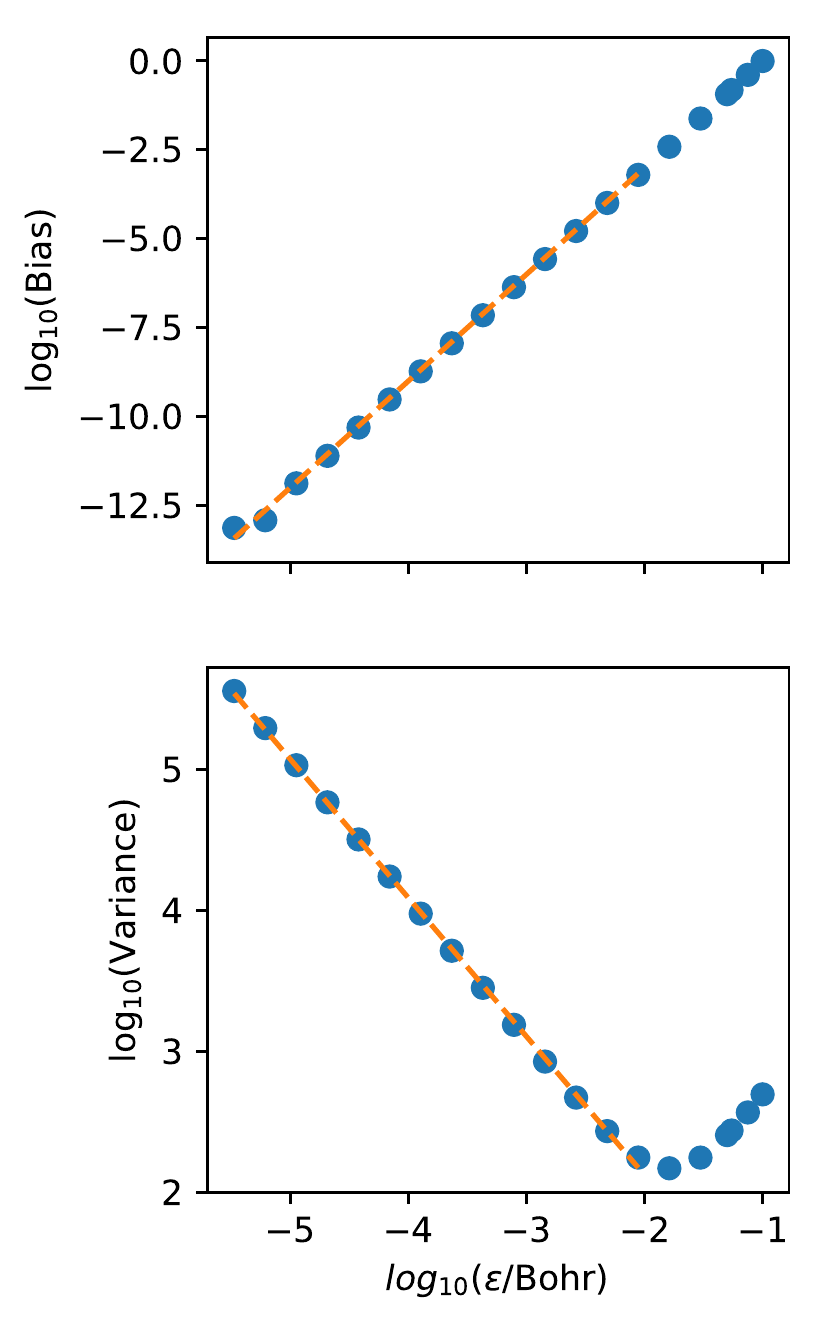}
\caption{Scaled bias and variance of $E_L\frac{\partial_p \Psi}{\Psi}$ evaluated by numerical integration from $l = -0.1$ to $l = 0.1$ across the node in Figure 1. The blue dots are the numerically integrated values and the orange curves indicate best fits to the functions $a\epsilon^3$ and $b + c/\epsilon$ for the bias and variance, respectively.}
\end{figure}

\begin{figure}
\includegraphics{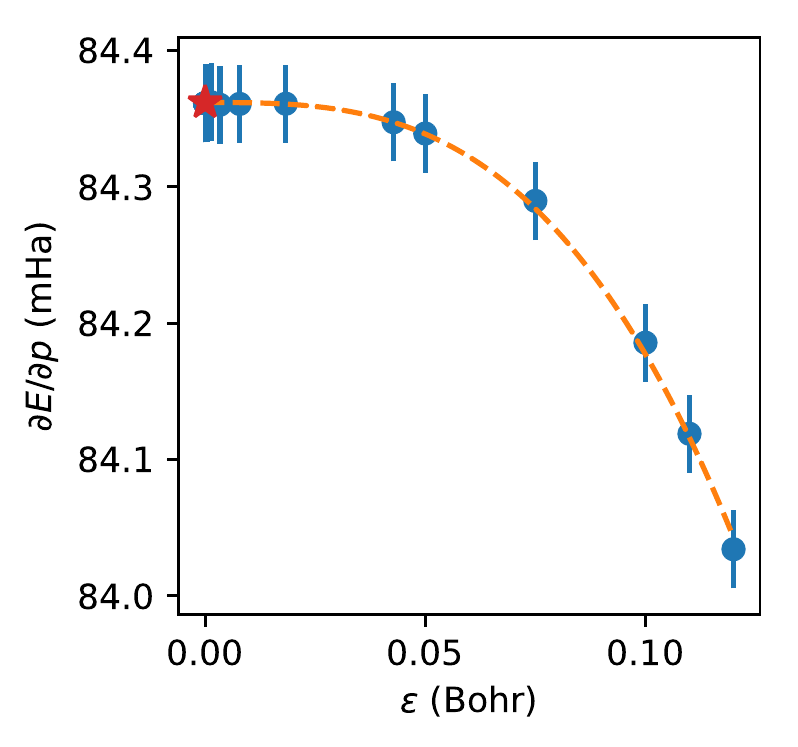}
\caption{Zero bias, finite variance extrapolation for $\partial E/\partial p$ using the regularized estimator. The blue points are evaluated using VMC, the orange curve is a fit to $a + b\epsilon^3$, and the red star denotes the extrapolated estimation. }
\end{figure}

We begin by verifying that the regularized integral in Eqn~\ref{eq:convergent_integral} has finite value across a node for $\epsilon > 0$. 
To do so, we evaluate the regularized integrand for various $\epsilon$ along a path which passes through a node $\vec{N}$, 
\begin{equation}
   \vec{R}(x) = \vec{N} + l \nabla \Psi(\vec{N})/|\nabla \Psi(\vec{N})|,
\end{equation}
 with $l \in [-0.1, 0.1]$ Bohr.
The results are shown in the lower plot of Figure 1.
For all values of $\epsilon$, the value of the integrand is pushed to zero at the node, removing the divergence in the integral.
The sharp increase in the integrand near $l=0$ as $\epsilon$ decreases is indicative of the increase in the value of the integral, and hence estimator variance, as $\epsilon \rightarrow 0$.
Shown in the upper plot is the first term in Eqn~\ref{eq:regularized_estimator} across the node, exhibiting no divergences.

By integrating across the normal coordinate $l$, the predicted $O(1/\epsilon)$, $O(\epsilon^3)$ scalings of the variance and bias are shown Figure 2.
The integration for both the bias and variance are carried out for the path in Figure 1 from $l = -0.1$ to $0.1$ Bohr and normalized by a factor $\int |\Psi|^2 dR$ along that path.
The predicted cubic bias is observed for six decades in $\epsilon$ while the $O(1/\epsilon)$ decrease in variance stands for four decades.
The increase in the variance after $\epsilon = 10^{-2}$ occurs due to the breakdown of the assumption $\Psi(\vec{R}) \simeq \nabla\Psi(\vec{N}) \cdot (\vec{R}-\vec{N})$, resulting in a linear increase in variance with $\epsilon$.
This change is not mirrored in the bias since the sub-leading order scaling with $\epsilon$ appears when the assumptions of $\hat{H}\Psi$ and $\partial_p \Psi \simeq const$ break down.

We conclude by carrying out the four step, zero-bias, finite variance extrapolation for $\partial E/\partial p$ proposed in the previous section, shown in Figure 3.
First, a standard VMC calculation with 200,000 steps and 2,000 configurations per step was carried out.
Then, $\partial E/\partial p$ was estimated using Eqn~\ref{eq:regularized_estimator} for $\epsilon$ between $1.2 \times 10^{-1}$ Bohr and $10^{-5}$ Bohr using the VMC configurations, shown in the blue data points.
Since the same configurations were used for each evaluation, the estimates have a strong statistical correlation and the predicted $\epsilon^3$ bias is clearly present, shown by the orange fit curve.
The intercept of the fit curve is shown by the red star and is the zero-bias estimate of $\partial E/\partial p$.
The variance can be deduced by the error bar of the blue data points, in this case 0.025 mHa.
Since the bias is zero within statistical errors for small values of $\epsilon$, most practical calculations can be carried out for a fixed value of $\epsilon$ between $10^{-5}$ and $10^{-2}$ Bohr without extrapolation.
Thus the regularized estimator is quite robust to $\epsilon$.

\section{Conclusion}
In this work, we derived and tested a simple regularized estimator for $\partial E/\partial p$ which has finite variance and can be extrapolated to zero bias.
The divergent variance present in the na\"ive Monte Carlo estimator is suppressed via multiplication by a polynomial function within a distance $\epsilon$ of the nodes of $\Psi(\vec{p}, \vec{R})$. 
We prove that the regularized estimator manages a finite variance by incurring a cubic bias which can be efficiently extrapolated to zero bias.
The extrapolation is carried out, and a finite variance, zero-bias estimate of $\partial E/\partial p$ is evaluated for a determinantal coefficient in a trial wave function for the LiH molecule.

The regularized estimator stands as an improvement to popular finite-variance estimation techniques for $\partial E/\partial p$.
These popular techniques require complex guiding \cite{Avella, Attaccalite2008, Zen2013}  or auxiliary wave functions \cite{Assaraf1999, doi:10.1063/1.1286598, Assaraf2003} in order to estimate $\partial E/\partial p$ for the trial wave function $\Psi(\vec{R}, \vec{p})$.
The regularized estimator does not require these additional, complicated wave functions, leading  to a simple algorithm for the finite variance, zero-bias, estimation of $\partial E/\partial p$ which requires only the trial wave function $\Psi(\vec{R}, \vec{p})$.

\begin{acknowledgments}
This research is part of the Blue Waters sustained-petascale computing project, which is supported by the National Science Foundation (awards OCI-0725070 and ACI-1238993) the State of Illinois, and as of December, 2019, the National Geospatial-Intelligence Agency. 
Blue Waters is a joint effort of the University of Illinois at Urbana-Champaign and its National Center for Supercomputing Applications. 
This material is based upon work supported by the U.S. Department of Energy, Office of Science, Office of Basic Energy Sciences, Computational Materials Sciences program under Award Number DE-SC-0020177.
\end{acknowledgments}

\bibliography{pgradregr}

\appendix*
\section{Constructing $f_\epsilon$}
The following are details on the construction of $f_\epsilon$ seen in Eqn~\ref{eq:regularizing_function}. 
We begin with the general form of a regularized estimator
$$
\hat{\theta_\epsilon} \equiv
\left\langle E_L(R) \frac{\partial_p \Psi(R)}{\Psi(R)} f_\epsilon(R) \right\rangle - \left\langle E_L(R) \right \rangle \left \langle \frac{\partial_p \Psi(R)}{\Psi(R)} f_\epsilon(R) \right\rangle
$$
where 
\begin{equation}
f_\epsilon(\vec{R}) = \begin{cases} 
      \sum_{n=1}^{\infty} a_n |\frac{\vec{x}}{\epsilon}|^n & |\frac{\vec{x}}{\epsilon}| < 1 \\
      1 & |\frac{\vec{x}}{\epsilon}| \ge 1 \\
   \end{cases},\ \vec{x} \equiv \frac{\Psi(\vec{R})\nabla \Psi(\vec{R})}{|\nabla \Psi(\vec{R})|^2}.
\end{equation} 
The previously divergent contribution to the estimator variance Eqn~\ref{eq:divergent_integral} is replaced by a finite integral
$$ \int_{|x/\epsilon|< 1} \Big(E_L\frac{\partial_p\Psi}{\Psi}\Big)^2 f_\epsilon^2 |\Psi|^2 dR,
$$
Since $|\vec{x}/\epsilon| \simeq |\nabla\Psi(\vec{N}) \cdot (\vec{R}-\vec{N})|/\epsilon|\nabla  \Psi(\vec{N})|$ as $|\vec{R} - \vec{N}| \rightarrow 0$, the integrand arbitrarily close to the nodal surface is a constant, meaning the integral is finite.

Unfortunately, the regularized estimator results in a linear-order biased estimation of $\partial E/\partial p$.
The bias results only from the first term and is written as 
$$
\text{Bias: } \int_{|\vec{x}/\epsilon|< 1} \Big(E_L\frac{\partial_p\Psi}{\Psi}\Big) (f_\epsilon - 1)|\Psi|^2 dR.
$$
As before, we take $\epsilon \rightarrow 0$ and Taylor expand the integrand to lowest order in $|\vec{R}-\vec{N}|$.
This yields an integrand, to lowest order in $\epsilon$, $a_1|\vec{x}/\epsilon| - 1$.
Carrying out the integration yields a bias that scales as $O(\epsilon)$ to lowest order in $\epsilon$.
The linear bias is inhibitive to zero bias extrapolation as the estimation bias is present for any value of $\epsilon$ while the variance increases rapidly with $\epsilon \rightarrow 0$.

A simple normalization condition on $f_\epsilon$ ensures a bias of $O(\epsilon^3)$, allowing for efficient extrapolation to zero bias.
The expression for the bias as $\epsilon \rightarrow 0$ after carrying out the necessary Taylor expansions is
$$
\lim_{\epsilon\rightarrow 0}\text{Bias} =  A \int_{|\vec{x}/\epsilon|< 1} (f_\epsilon - 1) dR.
$$
where $A$ is a constant resulting from $\hat{H}\Psi$ and $\partial_p \Psi \simeq const$ as $|\vec{R}-\vec{N}|\rightarrow 0$.
As such, the linear order bias can be removed by enforcing a normalization condition on $f_\epsilon$ on the domain of integration, 
$$
\lim_{\epsilon\rightarrow 0} \int_{|\vec{x}/\epsilon|< 1} (f_\epsilon - 1) dR = 0.
$$
The leading order contribution to the bias arises from the breakdown of the Taylor approximation $\hat{H}\Psi$ and $\partial_p \Psi \simeq const$, and results in a bias $O(\epsilon^3)$.

A final set of continuity and smoothness conditions on $f_\epsilon$ at $|\vec{x}/\epsilon| = 1$ reduce the magnitude of the cubic bias.
These two conditions can be written as 
$$
f(1) = 1, \nabla f(1) = 0.
$$
The three conditions, smoothness, continuity and normalization, can all be satisfied with a polynomial formula for $f$
$$
f_\epsilon(|\frac{\vec{x}}{\epsilon}|) = 7(\frac{\vec{x}}{\epsilon})^6 - 15(\frac{\vec{x}}{\epsilon})^4 + 9(\frac{\vec{x}}{\epsilon})^2.
$$
The normalization condition can be verified by noting the integration domain can be broken up as $\epsilon \rightarrow 0$
$$
\lim_{\epsilon\rightarrow 0} \int_{|\vec{x}/\epsilon|< 1}  dR  \rightarrow \int dN \int_{-\epsilon}^{\epsilon} dl(\vec{N})
$$
where the coordinate $\vec{N}$ is over the nodal surface, and the coordinate $l(\vec{N})$ is normal to that surface at $\vec{N}$.

\end{document}